\documentstyle[referee,graphicx]{mn}

\newif\ifAMStwofonts

\ifoldfss
  \ifCUPmtlplainloaded \else
    \NewTextAlphabet{textbfit} {cmbxti10} {}
    \NewTextAlphabet{textbfss} {cmssbx10} {}
    \NewMathAlphabet{mathbfit} {cmbxti10} {} 
    \NewMathAlphabet{mathbfss} {cmssbx10} {} 
  \fi
  \ifAMStwofonts
    \ifCUPmtlplainloaded \else
      \NewSymbolFont{upmath} {eurm10}
      \NewSymbolFont{AMSa} {msam10}
      \NewMathSymbol{\upi}     {0}{upmath}{19}
      \NewMathSymbol{\umu}     {0}{upmath}{16}
      \NewMathSymbol{\upartial}{0}{upmath}{40}
      \NewMathSymbol{\leqslant}{3}{AMSa}{36}
      \NewMathSymbol{\geqslant}{3}{AMSa}{3E}
      \let\oldle=3D\le     \let\oldleq=3D\leq
      \let\oldge=3D\ge     \let\oldgeq=3D\geq
      \let\leq=3D\leqslant \let\le=3D\leqslant
      \let\geq=3D\geqslant \let\ge=3D\geqslant
    \fi
  \fi
\fi 

\ifnfssone
  \newmathalphabet{\mathit}
  \addtoversion{normal}{\mathit}{cmr}{m}{it}
  \addtoversion{bold}{\mathit}{cmr}{bx}{it}
  \newmathalphabet{\mathbfit} 
  \addtoversion{normal}{\mathbfit}{cmr}{bx}{it}
  \addtoversion{bold}{\mathbfit}{cmr}{bx}{it}
  \newmathalphabet{\mathbfss} 
  \addtoversion{normal}{\mathbfss}{cmss}{bx}{n}
  \addtoversion{bold}{\mathbfss}{cmss}{bx}{n}
  \ifAMStwofonts
    \ifCUPmtlplainloaded \else
      \UseAMStwoboldmath
      \makeatletter
      \new@mathgroup\upmath@group
      \define@mathgroup\mv@normal\upmath@group{eur}{m}{n}
      \define@mathgroup\mv@bold\upmath@group{eur}{b}{n}
      \edef\UPM{\hexnumber\upmath@group}
      \new@mathgroup\amsa@group
      \define@mathgroup\mv@normal\amsa@group{msa}{m}{n}
      \define@mathgroup\mv@bold\amsa@group{msa}{m}{n}
      \edef\AMSa{\hexnumber\amsa@group}
      \makeatother
      \mathchardef\upi=3D"0\UPM19
      \mathchardef\umu=3D"0\UPM16
      \mathchardef\upartial=3D"0\UPM40
      \mathchardef\leqslant=3D"3\AMSa36
      \mathchardef\geqslant=3D"3\AMSa3E
      \let\oldle=3D\le     \let\oldleq=3D\leq
      \let\oldge=3D\ge     \let\oldgeq=3D\geq
      \let\leq=3D\leqslant \let\le=3D\leqslant
      \let\geq=3D\geqslant \let\ge=3D\geqslant
    \fi
  \fi
\fi 

\ifnfsstwo
  \DeclareMathAlphabet{\mathbfit}{OT1}{cmr}{bx}{it}
  \SetMathAlphabet\mathbfit{bold}{OT1}{cmr}{bx}{it}
  \DeclareMathAlphabet{\mathbfss}{OT1}{cmss}{bx}{n}
  \SetMathAlphabet\mathbfss{bold}{OT1}{cmss}{bx}{n}
  \ifAMStwofonts
    \ifCUPmtlplainloaded \else
      \DeclareSymbolFont{UPM}{U}{eur}{m}{n}
      \SetSymbolFont{UPM}{bold}{U}{eur}{b}{n}
      \DeclareSymbolFont{AMSa}{U}{msa}{m}{n}
      \DeclareMathSymbol{\upi}{0}{UPM}{"19}
      \DeclareMathSymbol{\umu}{0}{UPM}{"16}
      \DeclareMathSymbol{\upartial}{0}{UPM}{"40}
      \DeclareMathSymbol{\leqslant}{3}{AMSa}{"36}
      \DeclareMathSymbol{\geqslant}{3}{AMSa}{"3E}
      \let\oldle=3D\le     \let\oldleq=3D\leq
      \let\oldge=3D\ge     \let\oldgeq=3D\geq
      \let\leq=3D\leqslant \let\le=3D\leqslant
      \let\geq=3D\geqslant \let\ge=3D\geqslant
    \fi
  \fi
\fi 

\ifCUPmtlplainloaded \else
  \ifAMStwofonts \else 
    \def\upi{\pi}
    \def\umu{\mu}
    \def\upartial{\partial}
  \fi
\fi

\newcommand{\lsim}{\hbox{ \rlap{\raise 0.425ex\hbox{$<$}}\lower 0.65ex\hbox{$\sim$} }}
\newcommand{\gsim}{\hbox{ \rlap{\raise 0.425ex\hbox{$>$}}\lower 0.65ex\hbox{$\sim$} }}

\title[Slit Observations and Empirical Calculations for HII Regions]{Slit Observations and Empirical Calculations for HII Regions}
\author[I. F. Fernandes et al.]{I. F. Fernandes$^{1}$, R.
Gruenwald$^{2}$ and S. M. Viegas$^{3}$\\
$^{1, 2, 3}$Instituto de Astronomia, Geof\' \i sica e Ci\^encias
Atmosf\'ericas - USP, Rua do
Mat\~{a}o 1226, CEP 05508-900, S\~{a}o Paulo, Brazil\\
$^{1}$Laborat\'orio Nacional de Astrof\'isica, Rua Estados Unidos 154, CEP 37504-364 Minas Gerais, Brazil\\
$^{1}$E-mail:iran@astro.iag.usp.br\\
$^{2}$E-mail:ruth@astro.iag.usp.br\\
$^{3}$E-mail:viegas@usp.br\\
}

\begin{document}

\date{Accepted        .     Received        .}

\pagerange{\pageref{firstpage}--\pageref{lastpage}} \pubyear{2004}

\maketitle \label{firstpage}

\begin{abstract}
 When analysing HII regions, a possible source of systematic
error on empirically derived quantities, as  the gas temperature and the
chemical composition, is the limited  size of the slit used for
the observations. In order to evaluate this kind of systematic
error we use the photoionization code Aangaba  to create a
virtual photoionized region and mimic the effect of a slit
observation. A grid of models was built varying the ionizing
radiation spectrum emitted by a central stellar cluster, as well
as the gas abundance. The calculated line surface brightness was
then used to simulate slit observations and to derive empirical
parameters using the usual methods described in the literature.
Depending on the fraction  of the object covered
by the slit, the empirically derived physical parameters and
chemical composition can be different from those obtained from
observations of the whole object. This effect is mainly dependent
on the age of the ionizing stellar cluster. The low-ionization
lines, which originate in the outer layers of the ionized gas, are
more sensitive to the size of the area covered by the slit than
the high-ionization forbidden lines or recombination lines,
 since these lines are mainly produced closer to the inner radius
of the nebula. For a slit covering 50\% or less of the total area,
the measured [O III], [O II] and [O I] line intensities are less
than 78\%, 62\% and 58\% of  the total intensity for young HII
region (t $<$ 3 Myr); for older objects the effect due to the slit
 is less significant.
Regarding the temperature indicator T$_{[OIII]}$, the slit effects
are small ( usually less than  5\%) since this temperature is
derived from [OIII] high-ionization lines. On the other hand, for
the abundance (and temperature) indicator R$_{23}$, which depends
also on the [O II] line, the slit effect is slightly higher.
Therefore, the systematic error due to slit observations on the O
abundance is low, being usually less than 10\%, except for HII
regions powered by  stellar clusters with a relative low number of
ionizing photons between 13.6 and 54.4 eV, which create a smaller
O$^{++}$ emitting volume. In this case,  the systematic error on
the empirical O abundance deduced from slit observations is more
than 10\% when  the covered area is less than 50\%.

\end{abstract}


\section{Introduction}

For more than 30 years the empirical method for estimating
chemical abundances  from observed emission line intensities, as
proposed by Peimbert \& Costero (1969), has been largely used. The
method is simple enough to be applied to a wide variety of
objects, but can lead to some inconsistencies, as for example the
disagreement between abundances derived from forbidden and
recombination lines of C and O (Liu et al. 1995, Esteban et al.
2002). However, the method (with some improvements) is very
convenient for obtaining a first evaluation of the chemical
composition of large samples of objects as planetary nebulae, HII
regions and active galactic nuclei. The emitting gas is assumed to
have a high- and a low-ionized zones, characterized by the
temperature indicated, respectively, by the [O III] and [N II]
line intensity ratios. The electron density is usually derived
from the [S II] line ratio (see, for instance, Osterbrock 1989).
The various ions of the elements present in the gas are assumed to
be in either of these regions, following their ionization
potential.

As photoionization codes became friendly and available to users
(as the well-known Cloudy), several authors prefer to use them for
modeling the observed objects, in order to obtain the physical
conditions and the chemical composition of the emitting gas. The
calculations are usually performed assuming a spherically
symmetric object or a plan-parallel slab with a homogeneous
density distribution (see for instance P\'equignot 1986, Ferland
1995). These models could be considered as an improvement to the
empirical method described above.

Spectroscopic observations of HII regions in the Galaxy, as well
as in other galaxies, provide a large amount of data on
chemical composition (Edmunds \& Pagel 1984, van Zee et al. 1998,
Kobulnick et al. 1999) as well as on star formation rates
(Kennicutt 1992 and Izotov et al. 2004), which are the basis for a
better understanding of the chemical evolution of galaxies and of
the primordial nucleosynthesis. In the last decade the advent of
larger telescopes and modern detectors has increased the precision
of the data and raised the expectation for a better understanding of the
issues above. On other the hand, the physical parameters derived
from observational data of photoionized regions, using empirical
methods or one-dimension photoionization models, may introduce
systematic errors, decreasing the  possibility of reaching the
required precision. Some of these systematic errors, which reflect
on the elemental abundances, have already been discussed in the
literature (Steigman, Viegas \& Gruenwald 1997; Viegas, Gruenwald
\& Steigman 2000; Gruenwald, Steigman \& Viegas 2002; Martins \&
Viegas 2001, 2002). Regarding the star formation rate, a
comprehensive analysis of several indicators was recently carried
out by Rosa-Gonz\'alez, Terlevich \& Terlevich (2002).

Most of the emission-line data come from long-slit spectroscopic
observations of extended objects. Depending on the distance and on
the size of the object, the slit does not cover the whole emitting
region and the observed emission-line flux differs from the flux
produced by the whole object. Consequently, following the fraction
of the area of the nebula covered by the slit (hereafter referred
to as slit aperture), the emission line ratios may also differ.
Therefore, unless the extended object is far enough to be entirely
covered by the slit, the observed emission comes from a slab of
gas along the line of sight, with a cross-section equivalent to
the slit area projected on the object.  In this case, both the
empirical method and the photoionization simulations may lead to
erroneous results. In fact, the intensity of each line obtained
from slit observations come from a different fraction of its
emitting volume.

A comprehensive study of the variation of line strengths across
an HII region was carried out by Diaz et al. (1987) for NGC 604,
the largest H II region in the nearby galaxy M 33.
They present line intensities
relative to H$\beta$ in five different positions on the nebula.
As expected the values vary from point to point,
and the derived  O/H abundance may differ up to a factor of 2
(see their Table 1, positions A and E).

Another possible source of uncertainty is the effect of atmospheric
differential refraction as discussed by Filippenko (1982). In this work,
we  assume that the slit is in the optimal position.

As part of a program for searching for Wolf-Rayet stars (both WN
and WC) in galaxies, long-slit observations were performed for 14
galaxies (Fernandes et al. 2004). The main goal was to study the
relation of WR stars and the metallicity, as well as to compare
the results with the predictions of evolutionary synthesis models
(Schaerer \& Vacca 1998). Long-slit observations were performed at
the Palomar 200 inch telescope and the 3.6 m ESO/NTT, always using
a slit width of 1 arcsec.

Although these observations were aimed to look for the features
characterizing the presence of WR stars in the emission-line
spectra of star forming regions, the results provide a fair sample
of emission-lines that can be used to analyze most of the issues
presented above. Since the observed objects are nearby galaxies,
any study regarding chemical abundances as well as star formation
rates must account for the effect of the area covered by the slit.
The observed galaxies are located between 3 and 76 Mpc, except for
one galaxy, Mrk 309, at a distance of 170 Mpc. This implies that
the fraction of the area covered by the slit is less than 60\%,
except for the farthest galaxy. Bearing these values in mind,
we decided to evaluate the slit effect on the various quantities
derived by empirical methods.

Correction of the observational data by a slit aperture effect has
been largely discussed in the literature, usually associated to
the estimate of the galaxy luminosity (Kochanek, Pahre \& Falco
2001; Nakamura et al. 2003; Perez-Gonz\'{a}lez et al. 2003;
Brinchmann et al. 2004). Here, we analyze the effect of the slit
aperture on the empirical methods for chemical abundance
determination. A  description of how the results
from photoionization models can be used to simulate slit
observations are presented in \S 2.
The parameters derived from empirical methods are
discussed in \S 3 as a function of the fraction of the area
covered by the slit. Concluding remarks appear in \S 4.

\section{Photoionization Models and Slit Simulations}

In order to analyze the relationship between the fraction of area
covered by the slit and the parameters derived from empirical
calculations, we use photoionization models. We assume that the
physical conditions and the emission-line intensities obtained by
the numerical simulations correspond to the observed data from a
virtual nebula. These data are then used to derive parameters
by empirical methods, as
density and gas temperature, as well as the chemical composition.

The photoionization code Aangaba (Gruenwald \& Viegas 1992)
is used to simulate HII regions. In order to account for
the range of physical parameters found in HII regions, a grid of
models is built by varying the shape and intensity of the ionizing
radiation spectrum of the central stellar cluster, as well as the
gas chemical composition. Spherical symmetry is adopted for
calculating the diffuse radiation.

The shape of the ionizing spectrum is related to the age of the
young stellar cluster powering the nebula, while the intensity
(and the number of ionizing photons per second, Q$_H$), to its
initial mass. Stellar cluster ages of 0.0, 2.5, 3.3, 4.5 and 5.4
Myr (Cid-Fernandes et al. 1992) are used, with Q$_H$ in the range
10$^{49}$ to 10$^{53}$ photons s$^{-1}$. Depending on the stellar
cluster age, the corresponding mass spans from 150 to 10$^8$
M$_{\odot}$. The adopted chemical composition is in the range 0.1
to 1.5 Z$_{\odot}$, with solar values from Grevesse \& Anders
(1989) and  Grevesse \& Sauval (1998), keeping the solar ratio
between the heavy element abundances. The gas density is
 assumed constant and equal to 100 cm$^{-3}$.

The physical conditions are obtained for concentric shells with
increasing radii, starting at the inner radius of the nebula and
stopping when the gas is neutral (H$^+$/H $\leq$ 10$^{-2}$).  The
line-emission intensities are obtained adding the contribution
from each shell. The results at the gas outer radius correspond to
the total emission-line intensities produced by the virtual
nebula. The energy distribution of older clusters (3 $<$ t $<$ 6
Myr) is harder than that of younger clusters, leading to wider
recombination zones. Notice, however, that relative number
of photons between 13.6 and 54.4 eV for stellar clusters ages t
= 2.5 and 5.4 Myr is lower than for other ages.  It is then
expected that the behaviour of the ionic distribution of O$^+$ and
O$^{++}$ ions with the slit aperture for t = 5.4 Myr approaches
that of t = 2.5 Myr.

The code allows the calculation of line intensities at lines of
sight crossing the nebula and characterized by their projected
distance from the centre (see Gruenwald \& Viegas 1992 for
details). These results can then be used to obtain line
intensities coming from the region covered by the slit area
projected on the nebula, by integrating over the  specific studied
area. Recall that the virtual nebula is spherically symmetric.
Assuming that the slit is centred and its length is longer than
the nebula diameter, only the width of the slit limits the volume
of gas of each shell contributing to the line intensities. The
result for a slit with a projected width larger than the nebula
diameter must be equal to the line intensities coming from the
whole nebula, thus providing a test for our slit calculations when
compared with the results from the nebula simulations.

Once the emission-line intensities are obtained for a slit of a
given width, in the following characterized by the fraction of the
nebulae covered by the slit, the empirical methods are applied in
order to obtain the electron density and temperature of the gas,
as well as ionic and elemental abundances.

In order to derive empirical abundances from the observed
emission-lines, it is necessary to evaluate the gas temperature.
In the method first proposed by Peimbert \& Costero (1969), the
high-ionization region is characterized by the temperature
obtained from the [O III] $\lambda\lambda$4959,5007/$\lambda$4363
emission-line ratio, whereas the low-ionization
region by that from the
[N II] $\lambda\lambda$6548,6584/$\lambda$5754 emission-line
ratio. These line ratios depend on measured auroral line
intensities, usually weak and hard to measure mainly in high
abundance HII regions where the temperature is low. In this case,
another  temperature  indicator can be used as, for example,
R$_{23}$ = ([O III]$\lambda$5007 + [O II]$\lambda$3737)/H$\beta$
(Pagel et al. 1979) or R$_3$ = [O III]$\lambda$5007/H$\beta$
(Edmunds \& Pagel 1984).  Presently, many authors estimate the
chemical abundance of oxygen directly from the observed R$_{23}$
or R$_3$, using one of the calibrations available in the
literature (Edmunds \& Pagel 1984, Kobulnick et al. 1999, Pilyugin
2000, 2001 and  Kewley \& Dopita 2002). Regarding the density,
either the [S II] $\lambda$6717/$\lambda$6731 or the [O II]
$\lambda$3723/$\lambda$3730 emission-lines ratios are commonly
used as indicators. The former is more often used since the two [O
II] lines are usually blended.

Since O$^{++}$ and O$^{+}$ are the dominant ions in the HII
regions, the oxygen abundance can be obtained once the ionic
fractional abundances are derived from the observed lines.
Regarding other chemical elements like N and S, not all their ions
present in the ionized region are observable. Thus, ionization
correction factors (icfs), usually expressed in terms of the
abundance of O ions, must be used to access the elemental
abundances (Peimbert {\&} Torres-Peimbert 1977 and Peimbert {\&}
Torres-Peimbert 1987).


\section{Slit Empirical Results}

As mentioned above, a grid of HII region models was obtained
varying the age and the initial mass of the stellar cluster, as
well as the chemical composition of the gas. From the results, the
emission-line intensities for a virtual slit projected across the
nebula are obtained. The slit size is expressed by the fraction of
the total area covered by the slit, and varies from 0 to 1. In
order to discuss the effect of slit observations on the results
obtained from empirical methods, we first compare the behaviour of
measured  emission-line intensities of ions at different
ionization stages. Line ratios and derived quantities are then
analyzed.

\subsection{Emission-line intensities}

Because the low-ionization lines are produced in the outer zone of
the ionized region, their measured intensities are more sensitive
to slit aperture effects. Indeed, centred long-slit observations
cover a smaller fraction of the emitting volume of the low-ionization
lines than that of high-ionization lines,
 which are produced mainly in the inner zones.

In order to illustrate the effect of the slit aperture on the
measured line intensities, their ratios to the total line
intensity are shown in Figures 1 and 2 as a function of the
fraction of the total area covered by the slit. The recombination
lines H$\beta$ and HeII$\lambda$4686 are plotted in Fig. 1, while
the forbidden lines [O III]$\lambda$5007+4959, [O II]$\lambda$3727
and [O I]$\lambda$6300+6363 are in Figure 2. The results
correspond to models with ionizing stellar clusters of various
ages and Q$_H$ = 1.5$\times$ 10$^{49}$ s$^{-1}$, adopting gas
solar abundance.

The H$\beta$ line  is emitted throughout the ionized region and we
expect a mild dependence of the measured intensity on the slit
aperture. Notice however that the line intensity distribution
across the ionized region  depends  on the shape of the ionizing
radiation spectrum, so does the measured intensity for a given
slit aperture. As seen in Figure 1a, with a slit covering about
50\% of the total area, the measured H$\beta$ intensity is more
than 90\% of the total H$\beta$ emission for older HII regions,
while for younger clusters it is only 75\%, since these clusters
show a lack of photons with energy  higher than 24.6 eV. Regarding
the HeII line (Figure 1b), it is only produced by older HII
regions in the very inner ionized zone. In fact, even with a slit
covering only 30\% of the total area, the measured He II line
intensity is as high as 90\% of the total intensity. In HII
regions, the He$^+$ distribution across the photoionized gas is
very similar to that of H$^+$. Thus, the surface brightness
distribution of HeI recombination line is similar to that of
H$\beta$. For the recombination lines, the results shown in Figure
1 are roughly clustered around two curves depending on the
hardness of the ionizing spectrum.

\setcounter{figure}{0}
\begin{figure*}
\includegraphics[width=105mm,height=95mm]{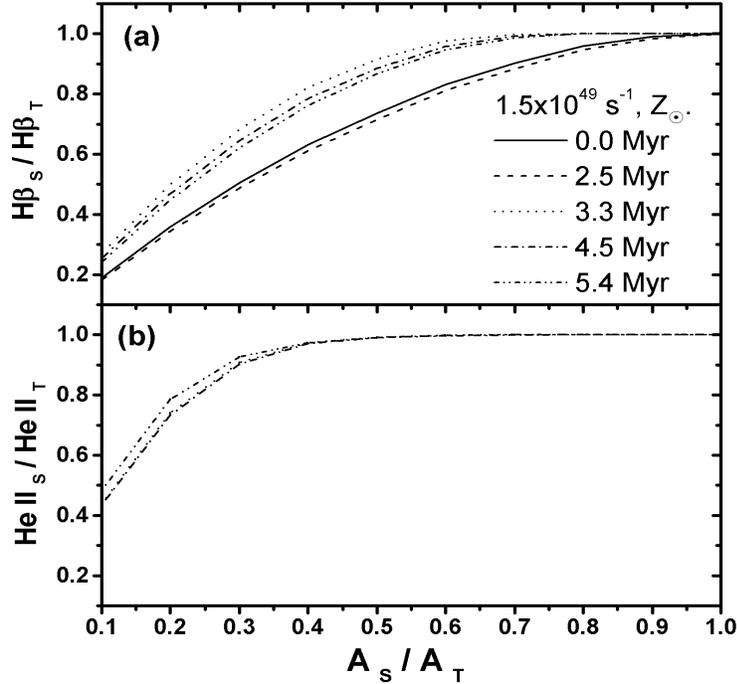}
\caption{Behaviour of (a) H$\beta\lambda$4861  and (b) He
II$\lambda$4686 line intensities with the area of the nebula
covered by the slit, A$_S$, normalized by the value corresponding to
the whole nebula, A$_T$. Both panels show results of models with Q$_H$
= 1.5$\times$ 10$^{49}$ s$^{-1}$ and solar abundance; different curves
correspond to different stellar ages.} \label{fig1}
\end{figure*}

\setcounter{figure}{1}
\begin{figure*}
\includegraphics[width=105mm,height=137mm]{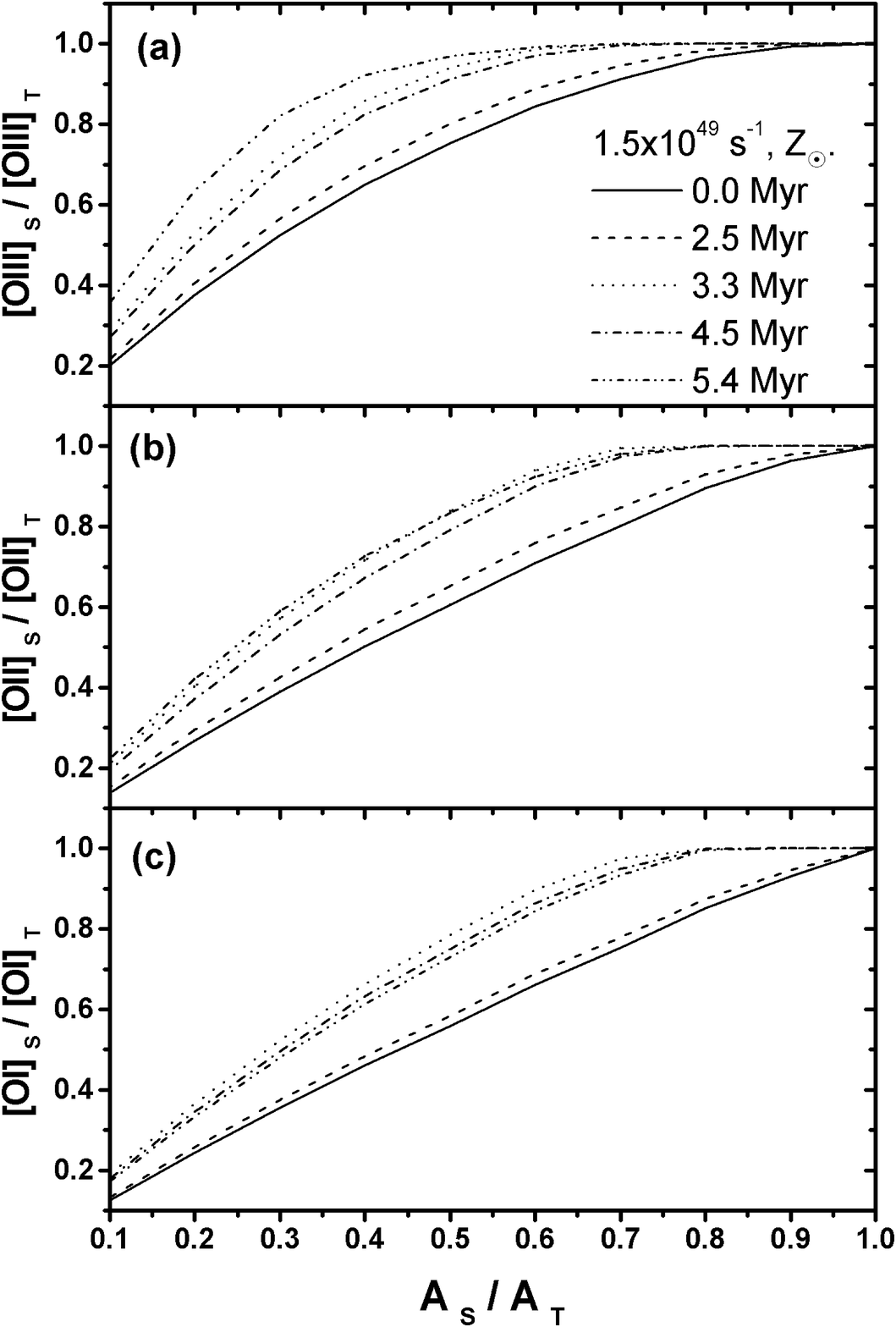}
\caption{Behaviour of (a) [O III]$\lambda$5007 , (b) [O
II]$\lambda$3727  and (c) [O I]$\lambda$6300  line intensities
with the area of the nebula covered by the slit. The notation is
the same as in Figure 1.} \label{fig2}
\end{figure*}

The dependence of the measured [O III], [O II] and [O I] line
intensities on the fraction of the area covered by the slit is
shown in Figures 2a, 2b and 2c, respectively. The O$^{++}$ ion
distribution across the ionized gas is similar to H$^+$. However,
forbidden lines are strongly dependent on the electron
temperature, so the results for [O III] line (Figure 2a) are more
dependent on the ionizing spectrum and are not as clustered as in
the recombination lines case. Regarding the low-ionization lines
(Figures 2b, 2c), which come mainly from the outer parts of the
nebula, we expect flatter gradients of the curves compared to
those of recombination and high-ionization lines. Indeed, we see
that, for a slit covering 50\% of the total area,  the measured
line intensity for [O III], [O II] and [O I] lines are about 78\%,
62\% and 58\% of the total intensity for an young HII region,
while it is 95\%, 82\% and 75\% for an older object.

Other low-ionization optical lines like [N II]$\lambda$6584+6548
and [S II]$\lambda$6716+6731, usually used to derive nitrogen and
sulphur abundances, show a dependence on the slit aperture
similar, respectively, to that of [O II] and [O I] lines.

For Q$_H$ higher than 1.5$\times$ 10$^{49}$ s$^{-1}$,  the slit
effect is more significant. For example, for a covered area of
50\%, the H$\beta$ line intensity is about 90\% of the total for
Q$_H$ = 1.5$\times$ 10$^{49}$ s$^{-1}$, while it can reach 70\%
and 55\% for Q$_H$ = 1.5$\times$ 10$^{51}$ s$^{-1}$ and
1.5$\times$ 10$^{53}$ s$^{-1}$, respectively. The same behaviour
is found for the high- and low-ionization forbidden lines.
However, for a given slit aperture, the variation of the line
intensity with Q$_H$ is smaller for low-ionization lines.

Models  with different chemical abundances show that the effect
of the slit aperture increases for increasing abundances. Indeed,
for a given ionizing spectra, the higher the chemical abundance
the farther from the central source is the zone where O$^{++}$
and O$^{+}$ are present and the higher their emission volumes.
Thus, in this case, the bulk of their lines comes from
the outer regions. It is then expected that
higher abundance nebulae are more affected by the slit size.

\subsection{Line intensity ratios and empirical temperature}

Temperature indicators, like T$_{[O III]}$ and R$_{23}$, as well
as empirical abundances are derived from line intensity ratios.
Since the line emitting volumes can be different, line ratios can
depend on the slit aperture.

The [O III]$\lambda$4363 is characteristic of high temperature
zones. Following the ionizing radiation spectrum, its emitting
volume may be more concentrated than that of [O III]$\lambda$5007.
In this case, we expect that their ratio may vary with the slit
aperture. Results for T$_{[O III]}$ as a function of the fraction
of the area covered by the slit are shown in Figure 3, for models
with Q$_H$ = 1.5$\times$ 10$^{49}$, solar abundances, and various
stellar cluster ages. It can be noticed that in the case of older
clusters (t $>$ 3 Myr), the T$_{[O III]}$ fraction is more
sensitive to the slit aperture mainly in the case of t = 5.4 Myr.
Compared to other old stellar cluster spectra, t = 5.4 Myr has a
relatively lower number of photons capable of ionizing H and He,
which are more efficient in heating the gas, creating a more
concentrated high temperature zone. In this case, only slits
covering more than 40\% of the total area will result on a
temperature value that differs less than 10\% from that obtained
from observations of the whole nebula. Younger stellar clusters
have a negligible number of photons which can produce O$^{++}$,
and the emitting volumes of the [O III] lines are similar. In this
case, the effect of the slit aperture is negligible. Regarding the
temperature indicator R$_{23}$, since it is mainly dominated by
the [O III]/H$\beta$ ratio, its behaviour as a function of the
slit aperture is very similar to that of T$_{[O III]}$,
although, due to the fact that it depends on the low-ionization
[O II] line, the systematic uncertainty due to the slit aperture
is slightly higher.

For models with Q$_H$ higher than those showed in Figure 3, the
slit effect tends to be negligible. Models with gas abundances
other than solar show that the slit effect on the intensity ratios
is similar or less significant than with solar values.

\setcounter{figure}{2}
\begin{figure*}
\includegraphics[width=105mm,height=59mm]{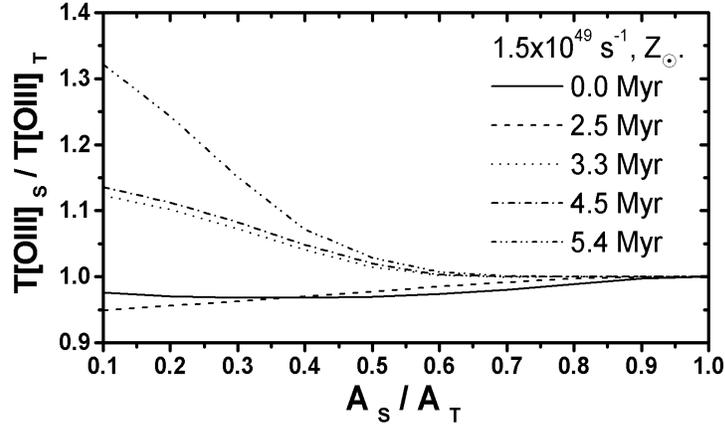}
\caption{Behaviour of empirical gas temperature, T$_{[OIII]}$,
with the area of the nebula covered by the slit. The notation  is
the same as in Figure 1.} \label{fig3}
\end{figure*}

\setcounter{figure}{3}
\begin{figure*}
\includegraphics[width=105mm,height=137mm]{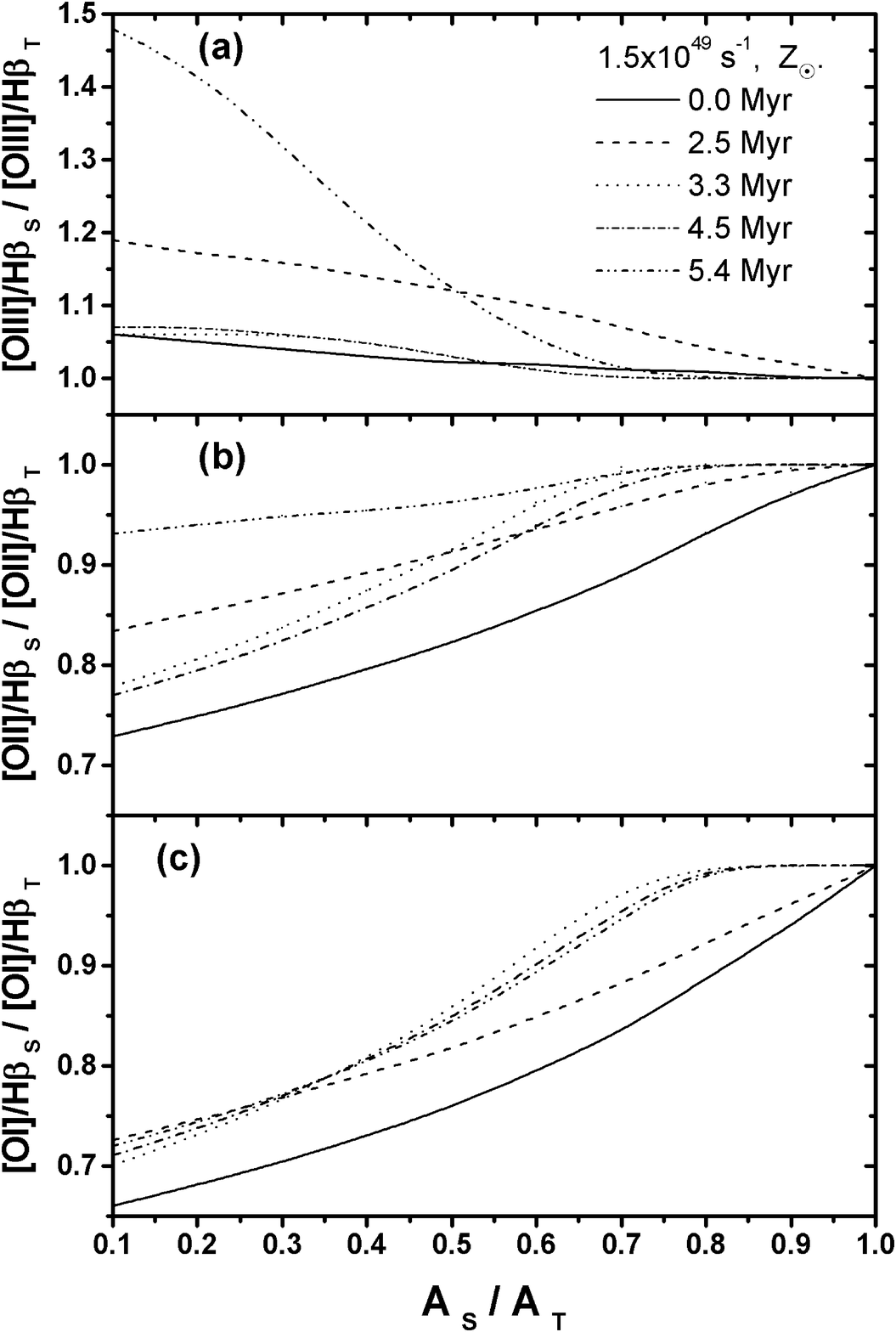}
\caption{Behaviour of line ratios (a) [O
III]$\lambda$5007/H$\beta\lambda$4861 , (b) [O
II]$\lambda$3727/H$\beta\lambda$4861  and (c) [O
I]$\lambda$6300/H$\beta\lambda$4861 with the area of the nebula
covered by the slit. The notation is the same as in Figure 1.}
\label{fig4}
\end{figure*}

To illustrate the behaviour of low and high ionization lines
relative to H$\beta$, the behaviour of O lines is shown in Figures
4a, 4b, and 4c. It can be seen that, as expected, the narrower the
slit the more underestimated are the [O I] and [O II] line ratios,
since they are produced in the outer zone of the HII region. On
the other hand, [O III] is usually overestimated because its
emission is more concentrated than that of H$\beta$ emission. The
effect is more significant for the two ionizing spectra showing
less photons with energy between  24.6 eV and 54.5 eV, i.e.,
stellar clusters of t = 2.4 and 5.4 Myr.

For a given stellar cluster age, the line ratios obtained from
models with higher Q$_H$ or different chemical composition (0.1 to
1.5 Z$_{\odot}$) differ less than 10\% from those presented in
Figures 4.

\subsection{Elemental Abundances}

It is well known that the oxygen abundance is tightly connected to
the gas temperature, since oxygen lines are efficient coolants of
the gas. It is then expected that the behaviour of the O/H
abundance ratio with the slit aperture is similar to that of
T$_{[O III]}$, since the [O III] lines are usually produced
throughout the nebula. However, the empirical O/H ratio depends
also on [O II]/H$\beta$, which is strongly dependent on the area
covered by the slit. The volume of the gas emitting [O II],
relative to the volume producing [O III], depends on the shape of
the ionizing radiation -- the larger the amount of high energy
photons the larger the recombination zone, thus the larger the
O$^+$ volume. The empirical O/H is much less dependent on the
value of Q$_H$, which characterizes the intensity of the ionizing
radiation, as well as on the adopted chemical composition (Figures
5a, 5b, and 5c).

Because of the similarities of the ionizing spectra for t = 2.5
and 5.4 Myr (see \S 2), the dependence of the empirical O
abundance on the area covered by the slit in these two cases is
similar (see Figure 5c). For HII regions ionized by such spectra,
the O empirical abundance derived from observations using a slit
partially covering the ionized region can be overestimated
relative to the abundance derived from observations of the whole
nebula. In the case of t = 2.5 Myr the systematic error is  about
12\% if the fraction of the covered area is less than 50\%, while
for t = 5.4 Myr this happens for a covered area less than 40\%
(see Figure 5c). For other cluster ages, the overestimation in the
abundance is  always less than 5\%.

Notice that if the R$_{23}$ indicator is used, the general
behaviour of O/H with the slit aperture is similar to that
described above. However, the dependence of R$_{23}$ on [O II]
increases the systematic uncertainty on O/H. Using the O/H -
R$_{23}$ calibration of Kobulnick et al. (1999), the uncertainty
is about 12\% for 2.5 Myr HII regions if the covered area is less
than 40\%, very similar to the result obtained with T$_{[OIII]}$.
However, for the case t = 5.4 Myr, the systematic uncertainty is
higher than 10\%, as long as the covered area is less than 50\%,
and linearly increases with decreasing area, reaching 58\% when
only 10\% of the nebula is covered by a centred slit.

\setcounter{figure}{4}
\begin{figure*}
\includegraphics[width=105mm,height=137mm]{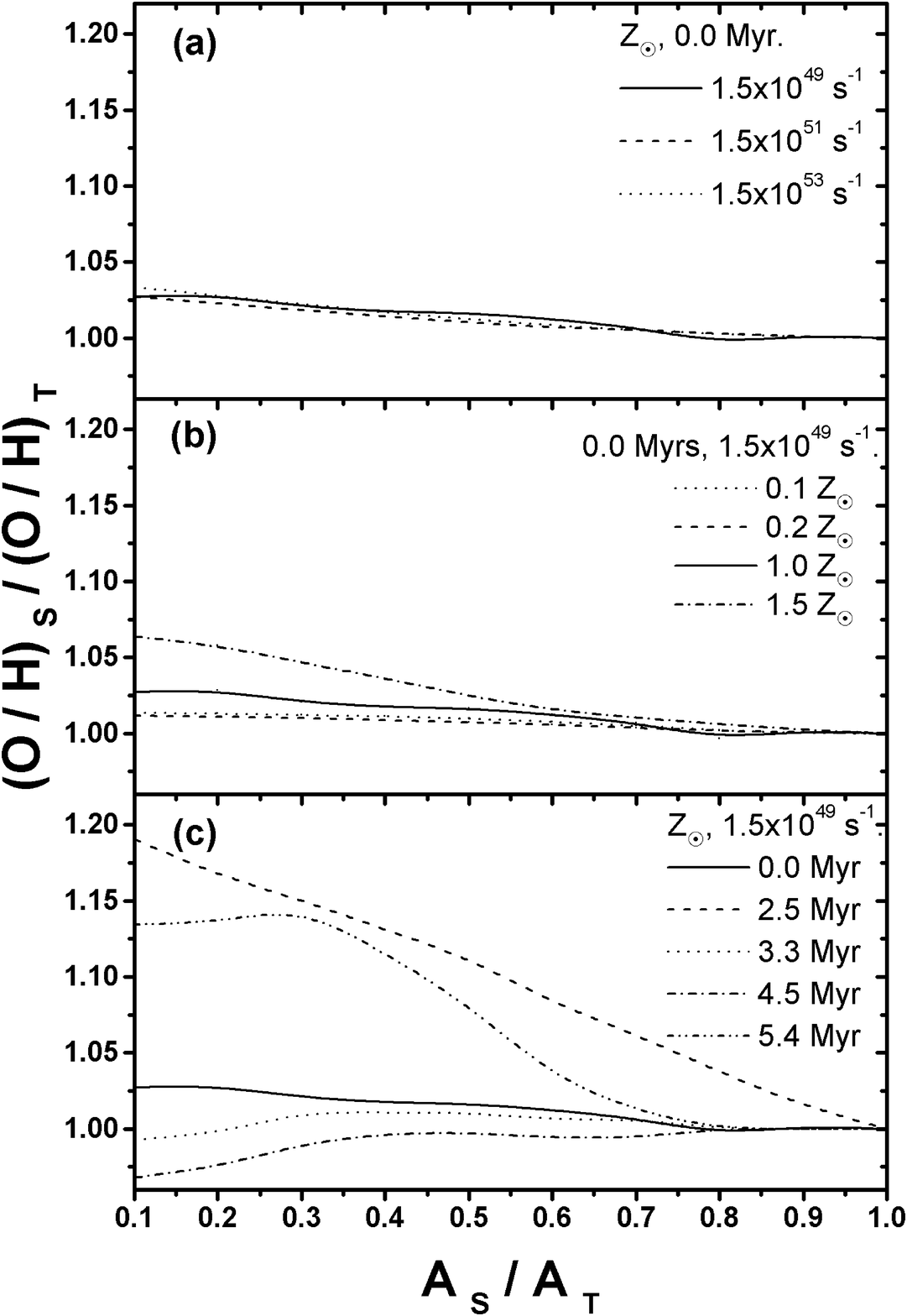}
\caption{Behaviour of the empirical oxygen abundance, O/H, with the
area of the nebula covered by the slit. Each panel shows results
of models varying Q$_H$, the elemental abundances, and the age of
the ionizing stellar cluster:
 (a) young stellar cluster (t = 0 Myr), solar abundance,
and three values for Q$_H$;
(b) young stellar cluster (t = 0 Myr),
Q$_H$ = 1.5$\times$ 10$^{49}$ s$^{-1}$ and elemental abundances
from 1.5 to 1/10 solar;
(c) Q$_H$ = 1.5$\times$ 10$^{49}$
s$^{-1}$, solar abundance and ionizing stellar clusters
of various ages from 0 to 5.4 Myr.} \label{fig5}
\end{figure*}

The derived He abundance usually varies less than 3\% with the
slit aperture (Figure 6a). Notice that the He abundance is derived
from recombination lines, mostly produced in the ionized gas.
Thus, the derived abundance is more sensitive to the shape of the
ionizing radiation spectrum than to the gas composition or the
ionizing radiation intensity adopted in the models.

Regarding the other elements with observable optical lines, their
abundances show a behaviour with the area covered by the slit
similar to that presented by O/H (Figures 6b and 6c), as long as
the abundance is obtained from low-ionization lines. Such
behaviour is expected for N/H and S/H since the measured lines are
low-ionization lines and their ionization correction factors are
estimated from the O ions. As for O/H, the systematic errors on
the empirical abundances derived for  N/H and S/H are less than
10\%, except for HII regions powered by stellar clusters with a
relative low number of photons between  13.6 and 54.4 eV (t = 2.5
and 5.4 Myr).

\setcounter{figure}{5}
\begin{figure*}
\includegraphics[width=105mm,height=137mm]{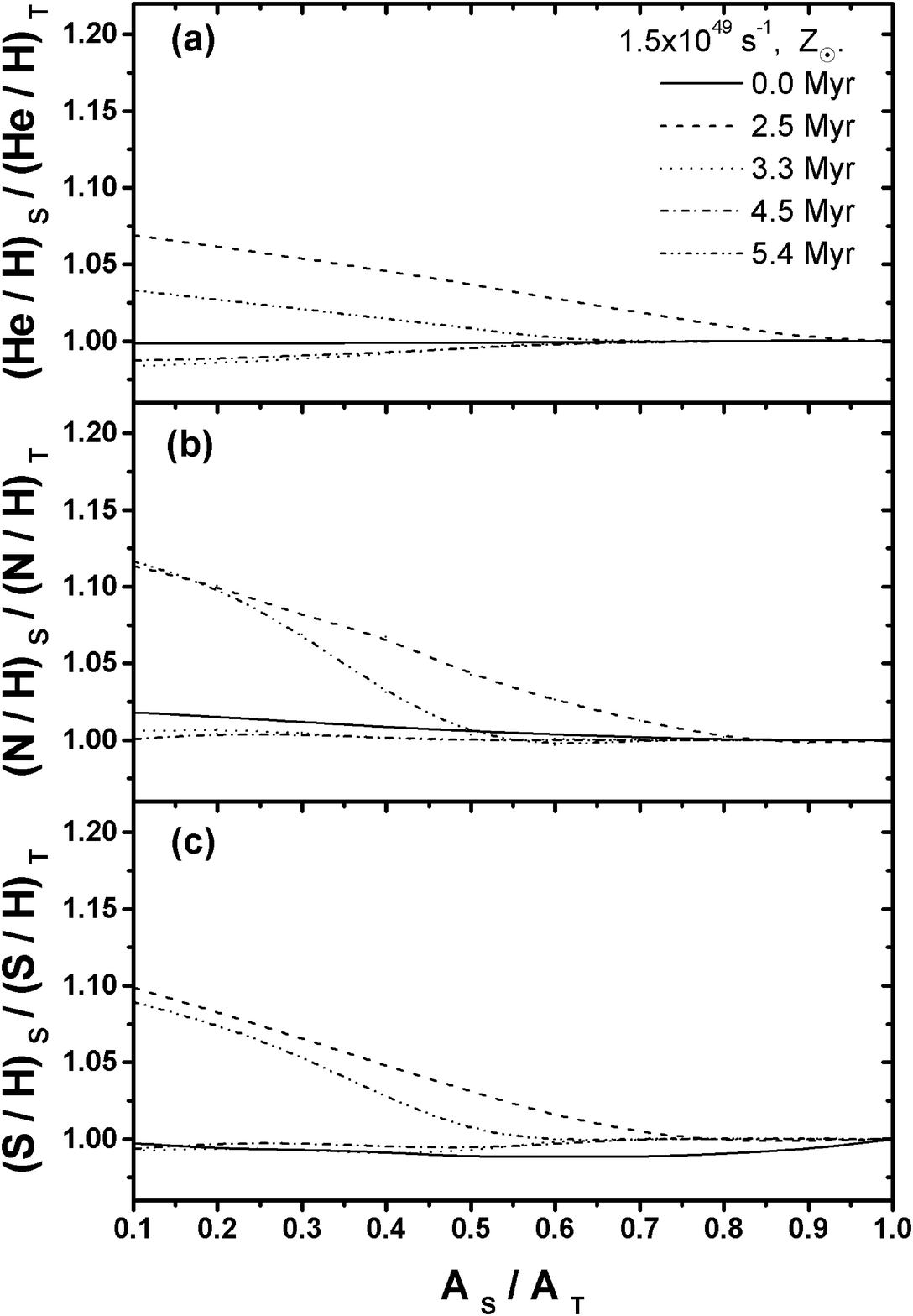}
\caption{Behaviour of (a) He/H , (b) N/H  and (c) S/H empirical
abundances with the area of the nebula covered by the slit. The
notation is the same as in Figure 1.} \label{fig6}
\end{figure*}

\vskip 3cm
\section{Concluding Remarks}

Because of the development of the instrumentation as well as  of
the observational techniques, the precision of the observational
data available in the last years has increased. Regarding the
emission-line regions present in galaxies, empirical methods to
derive the physical conditions and the elemental abundances of the
emitting gas are largely used. However, although useful, these
methods may introduce systematic errors which may be large than
the errors expected from the observations.

In this paper we have analyzed one possible source of systematic
error, which is the effect the slit size, i.e., the projected area
of the slit on the observed object. For this, we used a
photoionization code, Aangaba, to create a virtual HII region in
order to mimic and to analyze the effect of a slit on the
empirically derived properties of the nebula. The procedure
adopted in this work is equivalent to that of using a centred long
slit, with constant width, to observe the same kind of nebulae
located at different distances.

The main results are the following:

(a) The effect of the slit aperture on the empirically derived
physical parameters and on the chemical composition is mainly
dependent on the age of the ionizing stellar cluster. Because of
the characteristics of the spectral energy distribution, the
effect is more significant for t = 2.5 and 5.4 Myr.

(b) As expected, the low-ionization lines, which originate in the
outer layers of the ionized gas, are more sensitive to the size of
the area covered by the slit. Depending on the input parameters of
the photoionization model, [O II], [N II] and [S II]
emission-lines, relative to H$\beta$, may be underestimated by up to
30\% relative to the total emission.

(c) The HeI recombination line intensities, relative to H$\beta$,
are less sensitive to the slit width. The difference to the
emission of the whole nebula is less than 10\%, while for the HeII
lines the difference may reach 50 \%, since the He$^{++}$
distribution is more concentrated than that of H$^+$. For [O III]
lines, the effect is highly dependent on the ionizing spectrum.

(d) For the temperature indicators T$_{[OIII]}$ and R$_{23}$, the
overestimation can be more than 10\%, depending on the shape of
the ionizing spectrum. In particular for 5.4 Myr HII regions, the
systematic error on the O/H  is higher when the abundance is
derived from R$_{23}$, reaching 58\% for a centred slit covering
10\% of the nebula.

(e) The effect of the slit size on the empirical temperatures and
on the low-ionization emission-lines reflects on the elemental
abundances of O, N and S. For He abundance the effect of the slit
aperture is negligible.

In brief, when analyzing a large sample of data on star forming
regions using empirical methods, it is necessary to be aware of
possible systematic errors introduced by the slit effect. The
error is higher the smaller the area covered by the slit, tending
to be negligible ($<$ 10\%) when the fraction of the covered area
is larger than 50\%.  Remind that for a sample of 14 nearby
star-forming galaxies observed with a long slit with a width equal
to 1" (Fernandes et al. 2004), the fraction covered by the slit is
less than 60\%. Other samples of galaxies may also be affected by
systematic errors generated by slit observations, and could
reflect on the results regarding chemical evolution and star
formation rates derived from these samples.

It is also well known that extragalactic HII regions are largely
used to derive the primordial He abundance (see for instance,
Olive, Steigman \& Skillman 1997, Izotov \& Thuan 1998) and
references therein). These objects are usually distant and the
slit used in the observations may cover more than 50\% of the
area. However, a systematic error on the empirical gas temperature
leading to an error on the O/H empirical abundance may lead to a
systematic error on the primordial He abundance. This may be
significant enough to affect the conclusions drawn from the He
primordial abundance derivations (see for instance, Steigman et
al. 1997). A deeper analysis of the extragalactic HII region
sample is necessary to verify if the slit effect may release the
tension between the primordial He and D abundances (Olive,
Steigman, Walker 2000).

$Acknowledgements:$ This paper is partially supported by FAPESP
(00/06695-0), FAPESP (99/12721-5), CNPq (304077/77-1) and MCT/LNA
(381671/2005-4).


\bsp \label{lastpage}
\end{document}


\end{document}